\documentstyle[12pt] {article}
\newcommand{\beq}{\begin{equation}}
\newcommand{\eeq}{\end{equation}}
\newcommand{\beqa}{\begin{eqnarray}}
\newcommand{\eeqa}{\end{eqnarray}}
\newcommand{\ba}{\begin{array}}
\newcommand{\ea}{\end{array}}

\begin{document}

\begin{center}
{\large \bf Quantum Corrections to the Semiclassical \\
Quantization of the SU(3) Shell Model}\footnote{This work has been
partially supported by the Ministero della Universit\`a e della Ricerca
Scientifica e Tecnologica (MURST).}
\end{center}

\vskip 0.5 truecm

\begin{center}
{\bf V.R. Manfredi$^{(a)(b)}$ and L. Salasnich$^{(a)(c)}$}
\vskip 0.5 truecm
$^{(a)}$Dipartimento di Fisica "G. Galilei" dell'Universit\`a di Padova, \\
INFN, Sezione di Padova, \\
Via Marzolo 8, I 35131 Padova, Italy\footnote{Permanent address}\\
\vskip 0.5 truecm
$^{(b)}$Interdisciplinary Laboratory, SISSA, \\
Strada Costiera 11, I 34014 Trieste, Italy \\
\vskip 0.5 truecm
$^{(c)}$Departamento de Fisica Atomica, Molecular y Nuclear \\
Facultad de Ciencias Fisicas, Universidad "Complutense" de Madrid, \\
Ciudad Universitaria, E 28040 Madrid, Spain\\
\end{center}

\vskip 1.5 truecm
\begin{center}
Preprint DFPD/95/TH/09 \\
to be published in Modern Physics Letters B
\end{center}

\newpage

\begin{center}
{\bf Abstract}
\end{center}
\vskip 0.5 truecm
\par
We apply the canonical perturbation theory to the semi--quantal
hamiltonian of the SU(3) shell model.
Then, we use the Einstein--Brillowin--Keller quantization rule to obtain an
analytical semi--quantal formula for the energy levels, which is the
usual semi--classical one plus quantum corrections.
Finally, a test on the numerical accuracy of the semiclassical
approximation and of its quantum corrections is performed.

\vskip 1.5 truecm
\begin{center}
\end{center}

\newpage

\par
In the last few years, there has been considerable renewed interest in the
semi--classical approximation, due to the close connection to
the problem of the so--called quantum chaos [1,2]. One important aspect
is the semi--classical quantization formula of the energy
levels for quasi--integrable systems [3,4].
\par
It has recently been shown [5,6] that,
for perturbed non--resonant harmonic oscillators,
the algorithm of classical perturbation theory
may also be used in the quantum--mechanical perturbation theory,
with quantum corrections in powers of $\hbar$.
\par
In this paper, on the contrary, we calculate the quantum corrections to the
semi--classical quantization [3,4] of a many--body model
related to nuclear physics. Its classical counterpart,
obtained in the limit of the number of particles that goes to infinity,
is represented by a non--integrable hamiltonian
with two degrees of freedom [7,8,9,10].
The semi--classical quantization of this model has been studied in [7]
and here we calculate the quantum corrections and then analyze
their numerical accuracy.
\par
The model is a three--level schematic nuclear shell model,
whose hamiltonian is:
\beq
{\hat H} =\sum_{k=0}^{2}\epsilon_k {\hat G}_{kk}+{V\over 2}\sum_{k\neq l=0}^2
{\hat G}_{kl}^2,
\eeq
where
\beq
{\hat G}_{kl}=\sum_{m=1}^M {\hat a}_{km}^+ {\hat a}_{lm}
\eeq
are the generators of the SU(3) group. This model describes $M$ identical
particles in three, $M$--fold degenerate, single particle levels
$\epsilon_i$. There is a vanishing interaction
for particles in the same level and an equal interaction $V$ for particles in
different levels. We assume
$\epsilon_2=-\epsilon_0=\epsilon =1$, $\epsilon_1=0$.
\par
For the SU(3) model the semi--quantal hamiltonian [11] is defined as [8]:
\beq
H(p_1,p_2,q_1,q_2;M)=<q_1p_1,q_2p_2;M|{{\hat H}\over M}|q_1p_1,q_2p_2;M>,
\eeq
where $|q_1p_1,q_2p_2;M>$ is the coherent state, given by:
\beq
|q_1p_1,q_2p_2;M>=\exp{[z_1 G_{01} + z_2 G_{02}]} |00> ,
\eeq
with:
\beq
{1\over \sqrt{2M}}(q_k+ip_k)={z_k\over \sqrt{1+z_1^*z_1+z_2^*z_2}},
\;\;\; k=1,2
\eeq
and $|00>=\Pi_{k=1}^M a^+_{0k}|0>$ is the ground state.
Here $1/M$ plays the role of the Planck constant $\hbar$ [10].
\par
As discussed in great detail in [10], the semi--quantal
hamiltonian is:
$$
H(p_1,p_2,q_1,q_2;M) =
-1+{1 \over 2}(p_1^2+q_1^2)+(p_2^2+q_2^2)+
{1\over 4} \chi [1-{1\over M}] \times
$$
\beq
\times [(q_1^2+q_2^2)^2-
(p_1^2+p_2^2)^2-(q_1^2-p_1^2)(q_2^2-p_2^2)-4q_1q_2p_1p_2-
2(q_1^2+q_2^2-p_1^2-p_2^2)],
\eeq
with $\chi ={M V}/\epsilon$.
The phase space has been scaled to give $(q_1^2+q_2^2+p_1^2+p_2^2) \leq 2$.
The classical hamiltonian
can be obtained in the "thermodynamical" limit [10,12]:
\beq
H_{cl}(p_1,p_2,q_1,q_2)=\lim_{M\to \infty} H(p_1,p_2,q_1,q_2;M),
\eeq
and the semi--quantal hamiltonian is given by:
\beq
H(p_1,p_2,q_1,q_2;M)=H_{cl}(p_1,p_2,q_1,q_2)+H_{qc}(p_1,p_2,q_1,q_2;M),
\eeq
where $H_{qc}$ is the hamiltonian of quantum corrections.
\par
Through the canonical transformation in action--angle variables [11]:
\beq
q_k=\sqrt{2I_k}\cos{(\theta_k)}, \;\;\; p_k=\sqrt{2I_k}\sin{(\theta_k)},
\;\;\; k=1,2
\eeq
the semi--quantal hamiltonian can be written:
\beq
H(I_1,I_2,\theta_1,\theta_2;M)=H_0(I_1,I_2)+
\chi V(I_1,I_2,\theta_1,\theta_2;M),
\eeq
where:
\beq
H_0(I_1,I_2)=-1+I_1+2I_2,
\eeq
\beq
V(I_1,I_2,\theta_1,\theta_2;M)=[1-{1\over M}]
(1-I_1-I_2)[I_1\cos{(2\theta_1)}+I_2\cos{(2 \theta_2)}]+
I_1I_2\cos{(2\theta_2 - 2\theta_1)}.
\eeq
\par
We applied a canonical transformation
$(I_1,I_2,\theta_1,\theta_2) \to ({\tilde I}_1,{\tilde I}_2,
{\tilde \theta}_1,{\tilde \theta}_2)$ in order
to obtain a new hamiltonian that
depends only on the new action variables up to the second order in a
power series of $\chi$:
\beq
{\tilde H}({\tilde I}_1,{\tilde I}_2;M)=
{\tilde H}_0({\tilde I}_1,{\tilde I}_2)
+\chi {\tilde H}_1({\tilde I}_1,{\tilde I}_2;M)+
\chi^2 {\tilde H}_2({\tilde I}_1,{\tilde I}_2;M).
\eeq
It is well known that the canonical perturbation theory presents many
difficulties which are essentially related to the so--called small
denominators. The resonance of the unperturbed
frequencies $\omega_1 ={\partial H_0\over \partial I_1}=1,
\omega_2={\partial H_0\over \partial I_2}=2$:
\beq
m\omega_1+n\omega_2 =0,
\eeq
can lead to divergent expressions in the perturbative solution to the
problem. This drawback occurs only if the integer
numbers $m$ and $n$ are present
as Fourier harmonics in the perturbation theory.
We will show that the resonance condition (14) is not
satisfied up to the second order in $\chi$.
\par
We assume that the generator $S$ of the canonical transformation
may be expanded as a power series in $\chi$:
\beq
S({\tilde I}_1,{\tilde I}_2,\theta_1,\theta_2;M)=
{\tilde I}_1\theta_1 + {\tilde I}_2\theta_2
+\chi S_1({\tilde I}_1,{\tilde I}_2,\theta_1,\theta_2;M)
+\chi^2 S_2({\tilde I}_1,{\tilde I}_2,\theta_1,\theta_2;M).
\eeq
The generator $S$ satisfies the equations:
\beq
I_k={\partial S\over \partial \theta_k}={\tilde I}_k+
\chi {\partial S_1\over \partial \theta_k}
+\chi^2 {\partial S_2\over \partial \theta_k},
\eeq
\beq
{\tilde \theta}_k={\partial S\over \partial {\tilde I}_k}=\theta_k
+\chi {\partial S_1\over \partial {\tilde I}_k}
+\chi^2 {\partial S_2\over \partial {\tilde I}_k},
\eeq
with $k=1,2$. From the Hamilton--Jacobi equation:
\beq
H_0({\partial S\over \partial \theta_1},{\partial S\over \partial \theta_2})+
V({\partial S\over \partial \theta_1},{\partial S\over \partial
\theta_2}, \theta_1, \theta_2;M)=
{\tilde H}_0({\tilde I}_1,{\tilde I}_2)
+{\tilde H}_1({\tilde I}_1,{\tilde I}_2;M)
+{\tilde H}_2({\tilde I}_1,{\tilde I}_2;M),
\eeq
we have a number of differential equations obtained by equating the
coefficients of the powers of $\chi$:
\beq
{\tilde H}_0({\tilde I}_1,{\tilde I}_2)=
H_0({\tilde I}_1,{\tilde I}_2)=
-1+{\tilde I}_1+2{\tilde I}_2,
\eeq
\beq
{\tilde H}_1({\tilde I}_1,{\tilde I}_2;M)=
\left(\omega_1{\partial S_1\over \partial \theta_1} +
\omega_2{\partial S_1\over \partial \theta_2}
\right)+ V({\tilde I}_1,{\tilde I}_2, \theta_1, \theta_2;M),
\eeq
\beq
{\tilde H}_2({\tilde I}_1,{\tilde I}_2;M)=
\left( \omega_1{\partial S_2\over \partial \theta_1}
+\omega_2 {\partial S_2\over \partial \theta_2}
\right)
+ \left( {\partial V\over \partial I_1}{\partial S_1\over \partial \theta_1}
+{\partial V\over \partial I_2}{\partial S_1\over \partial \theta_2}
\right)
\eeq
The unknown functions ${\tilde H}_1$, $S_1$, ${\tilde H}_2$ and $S_2$
may be determined by averaging the time variation of the
unperturbed motion. At the first order in $\chi$ we obtain:
\beq
{\tilde H}_1({\tilde I}_1,{\tilde I}_2;M)=
{1\over 4\pi^2}\int_0^{2\pi}\int_0^{2\pi} d\theta_1 d\theta_2
V({\tilde I}_1,{\tilde I}_2, \theta_1, \theta_2;M) = 0,
\eeq
and
\beq
S_1({\tilde I}_1,{\tilde I}_2,\theta_1,\theta_2;M)=
-\sum_{\{(m,n)\}} {V_{mn}({\tilde I}_1,{\tilde I}_2;M)\over
(m\omega_1+n\omega_2)}\sin{(m\theta_1 +n\theta_2 )},
\eeq
where $\{(m,n)\}=\{(2,0),(0,2),(-2,2)\}$ are the Fourier harmonics
of the perturbation potential $V$. The resonance condition is not
satisfied, and we have:
$$
S_1({\tilde I}_1,{\tilde I}_2,\theta_1,\theta_2;M)=
-{1\over 2}[1-{1\over M}][(1-{\tilde I}_1
-{\tilde I}_2)({\tilde I}_1 \sin{(2 \theta_1)}+
$$
\beq
+{1\over 2}[1-{1\over M}]{\tilde I}_2 \sin{(2\theta_2)} )]-
{1\over 2}[1-{1\over M}]{\tilde I}_1
{\tilde I}_2 \sin{(2\theta_2 - 2\theta_1)}.
\eeq
At the second order in $\chi$:
$$
{\tilde H}_2({\tilde I}_1,{\tilde I}_2;M)=
{1\over 4\pi^2}\int_0^{2\pi}\int_0^{2\pi} d\theta_1 d\theta_2
\left( {\partial V\over \partial I_1}{\partial S_1\over \partial \theta_1}
+{\partial V\over \partial I_1}{\partial S_1\over \partial \theta_2}
\right)
$$
\beq
={1\over 4}[1-{1\over M}]
(-1+{\tilde I}_1+2{\tilde I}_2)(2{\tilde I}_1 -4{\tilde I}_1^2
+{\tilde I}_2-{\tilde I}_1{\tilde I}_2-{\tilde I}_2^2)
\eeq
and:
\beq
S_2({\tilde I}_1,{\tilde I}_2,\theta_1,\theta_2;M)=
-\sum_{\{(m,n)\}} {W_{mn}({\tilde I}_1,{\tilde I}_2;M)\over
(m\omega_1+n\omega_2)}\sin{(m\theta_1 +n\theta_2 )},
\eeq
where $\{(m,n)\}=\{(2,0),(4,0),(2,-4),(4,-4),(2,-2),(0,4),(2,2)\}$
are the Fourier harmonics of the function $W$, given by:
\beq
W({\tilde I}_1,{\tilde I}_2,\theta_1,\theta_2;M)=
{\tilde H}_2({\tilde I}_1,{\tilde I}_2;M)
-\left( {\partial V\over \partial I_1}{\partial S_1\over \partial \theta_1}
+{\partial V\over \partial I_2}{\partial S_1\over \partial \theta_2}
\right).
\eeq
In this case too, the resonance condition is not satisfied
and we have:
$$
S_2({\tilde I}_1,{\tilde I}_2,\theta_1,\theta_2;M)=
{1\over 8}[1-{1\over M}][3{\tilde I}_1{\tilde I}_2(1-{\tilde I}_1-{\tilde I}_2)
\sin{(2\theta_1)}+
$$
$$
+{\tilde I}_1(1-3{\tilde I}_1+2{\tilde I}_1^2-2{\tilde I}_2+3{\tilde I}_1
{\tilde I}_2+{\tilde I}_2^2)\sin{(4\theta_1)} +
$$
$$
+3{\tilde I}_1{\tilde I}_2({\tilde I}_1+{\tilde I}_2^2-1)
\sin{(2\theta_1-4\theta_2)}+
$$
$$
+{\tilde I}_1{\tilde I}_2({\tilde I}_2-{\tilde
I}_1)\sin{(4\theta_1-4\theta_2)}+
$$
$$
+3{\tilde I}_1{\tilde I}_2(1-{\tilde I}_1-{\tilde I}_2)
\sin{(2\theta_1-2\theta_2)} +
$$
$$
+{1\over 4}{\tilde I}_2(1-2{\tilde I}_1+{\tilde I}_1^2-3{\tilde I}_2+
3{\tilde I}_1{\tilde I}_2+2{\tilde I}_2^2)\sin{(4\theta_2)}+
$$
\beq
+3{\tilde I}_1{\tilde I}_2({\tilde I}_1+{\tilde I}_2-1)
\sin{(2\theta_1+2\theta_2)} ]
\eeq
\par
In conclusion:
\beq
{\tilde H}({\tilde I}_1,{\tilde I}_2;M)=
-1+{\tilde I}_1+2{\tilde I}_2 +
{\chi^2 \over 4}[1-{1\over M}]
(-1+{\tilde I}_1+2{\tilde I}_2)(2{\tilde I}_1 -4{\tilde I}_1^2
+{\tilde I}_2-{\tilde I}_1{\tilde I}_2-{\tilde I}_2^2).
\eeq
This approximate semi--quantal hamiltonian depends only on the actions.
Thus, a semi--quantal quantization formula may be obtained by applying
the Einstein--Brillowin--Keller rule [2,3]:
\beq
{\tilde I}_{k}=(n_k+{1\over 2}){1\over M}, \;\;\; k=1,2
\eeq
where $1/M$ plays the role of the Planck constant $\hbar$.
In this way we have:
\beq
E_{n_1 n_2}(M)=E_{n_1 n_2}^{sc}(M)+E_{n_1 n_2}^{qc}(M)
\eeq
where:
$$
E_{n_1 n_2}^{sc}(M)= -1+(n_1+{1\over 2}){1\over M} +2(n_2+{1\over 2}){1\over M}
+{\chi^2\over 4}[-1+(n_1+{1\over 2}){1\over M}
+2(n_2+{1\over 2}){1\over M} ]\times
$$
\beq
\times [2(n_1+{1\over 2}){1\over M}  -4(n_1+{1\over 2})^2{1\over M^2}
+(n_2+{1\over 2}){1\over M} -(n_1+{1\over 2})(n_2+{1\over 2})
{1\over M^2}-(n_2+{1\over 2})^2{1\over M^2}],
\eeq
is the semi--classical quantization formula, and
$$
E_{n_1 n_2}^{qc}(M)= -{\chi^2\over 4M}[-1+(n_1+{1\over 2}){1\over M}
+2(n_2+{1\over 2}){1\over M} ]\times
$$
\beq
\times [2(n_1+{1\over 2}){1\over M}  -4(n_1+{1\over 2})^2{1\over M^2}
+(n_2+{1\over 2}){1\over M} -(n_1+{1\over 2})(n_2+{1\over 2})
{1\over M^2}-(n_2+{1\over 2})^2{1\over M^2}],
\eeq
are the quantum corrections.
\par
In order to test the accuracy of the semiclassical approximation and
its quantum corrections, the eigenvalues of the hamiltonian (1) must
be calculated. A natural basis can be written:
$|bc>$, meaning b particles in the second level, c in the third and, of
course, $M-b-c$ in the first level. In this way $|00>$ is the ground state
with all the particles in the lowest level [7,8]. We can write the general
basis state:
\beq
|bc>=\sqrt{1\over b!c!}{\hat G}_{21}^{b}{\hat G}_{31}^{c}|00>,
\eeq
where $\sqrt{1\over b!c!}$ is the normalizing constant.
\par
We can calculate the expectation values of
${\hat H}\over M$ and, therefore, the eigenvalues and eigenstates of
${\hat H}\over M$.
In this way, the energy spectrum range is independent of the
number of the particles:
\beq
<b^{'}c^{'}|{{\hat H}\over M}|bc>={1\over M}(-M+b+2c)\delta_{bb^{'}}
\delta_{cc^{'}}-{\chi \over 2M^{2}}Q_{b{'}c^{'},bc},
\eeq
where:
$$
Q_{b{'}c^{'},bc}=\sqrt{b(b-1)(M-b-c+1)(M-b-c+2)}\delta_{b-2,b^{'}}
\delta_{cc^{'}}
$$
$$
+\sqrt{(b+1)(b+2)(M-b-c)(M-b-c-1)}\delta_{b+2,b^{'}}\delta_{cc^{'}}
$$
$$
+\sqrt{c(c-1)(M-b-c+1)(M-b-c+2)}\delta_{b,b^{'}}\delta_{c-2,c^{'}}
$$
$$
+\sqrt{(c+1)(c+2)(M-b-c)(M-b-c-1)}\delta_{b,b^{'}}\delta_{c+2,c^{'}}
$$
$$
+\sqrt{(b+1)(b+2)c(c-1)}\delta_{b+2,b^{'}}\delta_{c-2,c^{'}}
$$
\beq
+\sqrt{b(b-1)(c+1)(c+2)}\delta_{b-2,b^{'}}\delta_{c+2,c^{'}}
\eeq
and $\chi =MV/\epsilon$.
The expectation value $<{{\hat H}\over M}>$ is real and
symmetric. For a given number
of particles M, we can set up the complete basis state, write down the matrix
elements of $<{{\hat H}\over M}>$ and then diagonalize $<{H\over M}>$
to find its eigenvalues.
$<{H\over M}>$ connects only states with $\Delta b=-2,0,2$ and
$\Delta c=-2,0,2$ which makes the problem easier. We group states with b,c
even; b,c odd; b even and c odd; b odd and c even. This means that
$<{{\hat H}\over M}>$ becomes block diagonal containing 4 blocks which can be
diagonalized separately. These matrices are referred to as ee, oo, oe and eo
(for further details see also [17]).
\par
Then we compare these "exact" levels to those obtained by the
semi--quantal perturbation theory.
A very good agreement is displayed (see Fig. 1).
\par
In Table 1, we show the difference between the "exact" levels
and those obtained by the semi--classical and semi--quantal
perturbation theory.
We observe that the algorithm provided
by the semi--quantal perturbation theory gives better
results than that of the ordinary semi--classical
perturbation theory.
\par
Obviously if $1/M$, no matter how small, is kept fixed, this
semi--quantal approximation on the individual levels has
the meaning of a perturbation theory in $1/M$ [5,6,13].
Therefore, the accuracy of the approximation decreases for
higher levels [14].
To obtain a better agreement it is necessary, as is well known, to
implement the classical limit $1/M \to 0$, $n_k \to \infty$
and, at the same time, to keep the action ${\tilde I_k}=(n_k+1/2)/M$
constant [15,16].
\par
Finally, we stress that, for systems with a finite number of Fourier
harmonics, like the SU(3) model, rational frequencies do not give rise
to the problem of small denominators up to a certain order of
the canonical perturbation theory.

\newpage

\begin{center}
{\bf Acknowledgments}
\end{center}
\vskip 0.5 truecm
\par
The authors are greatly indebted to Prof. S. Graffi and Prof. G. Alvarez
for many enlightening discussions. One of us (L.S.) is also indebted to
Prof. J.M.G. Gomez for his kind hospitality at the Departamento
de Fisica Atomica, Molecular y Nuclear de Universidad Complutense de Madrid.

\newpage

\parindent=0.pt

\section*{References}
\vspace{0.6 cm}

[1] {\it Chaos and Quantum Physics}, in Les Houches Summer
School, Course LII, 1989, Ed. By M.J. Giannoni, A. Voros and J.
Zinn-Justin, Elsevier Science Publishing (1989)

[2] A.M. Ozorio de Almeida, {\it Hamiltonian Systems: Chaos and
Quantization}, Cambridge University Press (1990);
M.C. Gutzwiller, {\it Chaos in Classical and Quantum Mechanics},
Springer--Verlag (1990);
{\it From Classical to Quantum Chaos}, SIF Conference Proceedings, vol.
{\bf 41}, Ed. G. F. Dell'Antonio, S.
Fantoni, V. R. Manfredi (Editrice Compositori, Bologna, 1993)

[3] V.P. Maslov and M.V. Fedoriuk, {\it Semi--Classical Approximation in
Quantum Mechanics}, Reidel Publishing Company (1981)

[4] A.R.P. Rau, Rev. Mod. Phys. {\bf 64} (1992) 623;
P.A. Braun, Rev. Mod. Phys. {\bf 65} (1993) 115

[5] S. Graffi, T. Paul: Commu. Math. Phys. {\bf 107} (1987) 25

[6] S. Graffi, M. Degli Esposti, Herczynski: Ann. Phys. (N.Y.) {\bf 208}
(1991) 364

[7] R. Williams and S. Koonin, Nucl. Phys. A {\bf 391} (1982) 72

[8] D. Meredith, S. Koonin and M. Zirnbauer, Phys. Rev. A {\bf 37} (1988)
3499

[9] P. Leboeuf and M. Saraceno, Phys. Rev. A {\bf 41} (1990) 4614

[10] W.M. Zhang, D.H. Feng: Phys. Rev. A {\bf 43} (1991) 1127

[11] M. Born, {\it The Mechanics of the Atom}, G. Bell and Sons Ltd.,
London (1927); W. Dittrich, M. Reuter, {\it Classical and Quantum
Dynamics}, Springer--Verlag (1992)

[12] W.M. Zhang, D.H. Feng, R. Gilmore: Rev. Mod. Phys. {\bf 62} (1990) 867;
W.M. Zhang, D.H. Feng, J.M. Yuan: Phys. Rev. A {\bf 42} (1990) 7125

[13] S. Graffi: in {\it Probabilistic Methods in Mathematical Physics}, Ed.
F. Guerra, M.I. Loffredo and C. Marchioro (World Scientific, 1992)

[14] T. Prosen and M. Robnik: J. Phys. A {\bf 26} (1993) L37

[15] G. Alvarez, S. Graffi, H.J. Silverstone: Phys. Rev. A {\bf 38}
(1988) 1687;

[16] S. Graffi, V.R. Manfredi, L. Salasnich: Nuovo Cim. B {\bf 109}
(1994) 1147

[17] V.R. Manfredi, L. Salasnich, L. Dematte: Phys. Rev. E {\bf 47}
(1993) 4556

\newpage

\parindent=0.pt
\section*{Table Captions}
\vspace{0.6 cm}

{\bf Table 1}: The differences for the first 10 levels,
with $\chi =0.75$ and $M=100$, for the $eo$ class.
$E^{ex}$ are the "exact" levels, $E^{sc}$ are the semi--classical levels,
and $E^{sq}$ are the semi--quantal levels.

\newpage

\parindent=0.pt
\section*{Figure Captions}
\vspace{0.6 cm}

{\bf Figure 1}: Comparison between "exact" levels (left)
and those obtained by the semi--quantal perturbation theory (right);
with $\chi =0.75$ and $M=100$ for the $eo$ class.

\newpage

\begin{center}
\begin{tabular}{|cc|} \hline\hline
    $|E^{ex}-E^{sc}|$ & $|E^{ex}-E^{sq}|$  \\ \hline
  1.3086796$\cdot 10^{-3}$ & 1.1860132$\cdot 10^{-3}$ \\
  1.6146898$\cdot 10^{-3}$ & 1.3964772$\cdot 10^{-3}$ \\
  2.7745962$\cdot 10^{-4}$ & 2.2178888$\cdot 10^{-4}$ \\
  1.4380813$\cdot 10^{-3}$ & 1.1500716$\cdot 10^{-3}$ \\
  1.5463829$\cdot 10^{-3}$ & 1.3815761$\cdot 10^{-3}$ \\
  1.0503531$\cdot 10^{-3}$ & 7.1579218$\cdot 10^{-4}$ \\
  1.9035935$\cdot 10^{-3}$ & 1.6558170$\cdot 10^{-3}$ \\
  4.4906139$\cdot 10^{-4}$ & 3.5130978$\cdot 10^{-4}$ \\
  6.0987473$\cdot 10^{-4}$ & 2.4980307$\cdot 10^{-4}$ \\
  1.8021464$\cdot 10^{-3}$ & 1.4950633$\cdot 10^{-3}$ \\ \hline\hline
\end{tabular}
\end{center}

\vskip 0.8 truecm
{\bf Table 1}

\end{document}